\documentclass[conference]{IEEEtran}

\ifCLASSINFOpdf
\else
\fi
\usepackage{graphicx}
\usepackage{soul}
\begin{document}

\title{LoRaWAN attack in military use case}

\author{\IEEEauthorblockN{Georges Derache, Mounira Msahli, Aurélien Botbol, Fabien Romain, Jérôme Champlon, Gauthier Canet, }
\IEEEauthorblockA{Télécom Paris, Institut Polytechnique de Paris}
\IEEEauthorblockA{(georges.derache, mounira.msahli, aurelien.botbol, fabien.romain, jerome.champlon, gauthier.canet)}@telecom-paris.fr\\}

\maketitle

\begin{abstract}
The importance of the development of IoT and LoRaWAN in military applications has been widely established. Since security is one of its important challenges, in this paper we study two attacks scenarios: replay and sniff attacks on military LoRaWAN network. The aim is to highlight cybersecurity threats that must be taken into consideration when using such technology in critical context.

\end{abstract}

\IEEEpeerreviewmaketitle

\section{Introduction}


We have observed the increasing development of the IoT (Internet of Things) in several diverse fields. The use of these technologies is booming. Today, approximately forty billion devices, with an estimated seventy-five billion connected objects by 2025 \cite{1}. This increase use of IoT technology is mainly the result of the development of intelligent automation in homes, the growing interconnection of user services with businesses, and the emergence of smart cities. \\

In fact, its long-distance communications capabilities and excellent energy autonomy are a great assets that encourage its use in different domains. Like the civilian world, the military sector is actively seeking to develop the use of IoT to adapt and improve its combat systems. In this paper, we consider the use of LoRaWAN technology in national defense sector. This includes systems for locating soldiers and monitoring health using the LoRaWAN model \cite{2}, or a LoRa-based UAV (Unmanned Aerial Vehicle) fingerprinting framework \cite{3}, or even the development of the TRACE project, which aims to automate the maintenance of ammunition using a tool for remote reading and LoRaWAN transmission of storage conditions \cite{4}. One of the common points of indicated military applications is the transit of sensitive information to which an armed force should have access in all circumstances, without revealing any secrets to the enemy. \\
 
In a world where cyberwarfare is omnipresent \cite{5}, the enemy might want to harm its opponent by attacking LoRaWAN transmission systems. The aim of this paper is to study the military's threats by giving some attack scenarios of LoRaWAN network. In this paper, we consider sniffing and replay attacks.\\
 
The rest of this paper is organized as follows: Section II describes the military scenarios, Section III presents the implementation and results of our solution, and Section IV provides a conclusion.\\


\section{Proposed approach and mechanism}


In this paper, we consider two scenarios of attacks on military LoRaWAN networks (see Fig. 2). We consider a network for monitoring shell containers on the front line with GNSS position feedback based on LoRaWAN technology. The aim here is to detect, sniff this network, intercept it, exploit its metadata (S1: scenario 1), and then neutralize the link between the device and the gateway (S2: scenario 2). \par

\begin{figure}[htp]
    \centering
    \includegraphics[width=8cm]{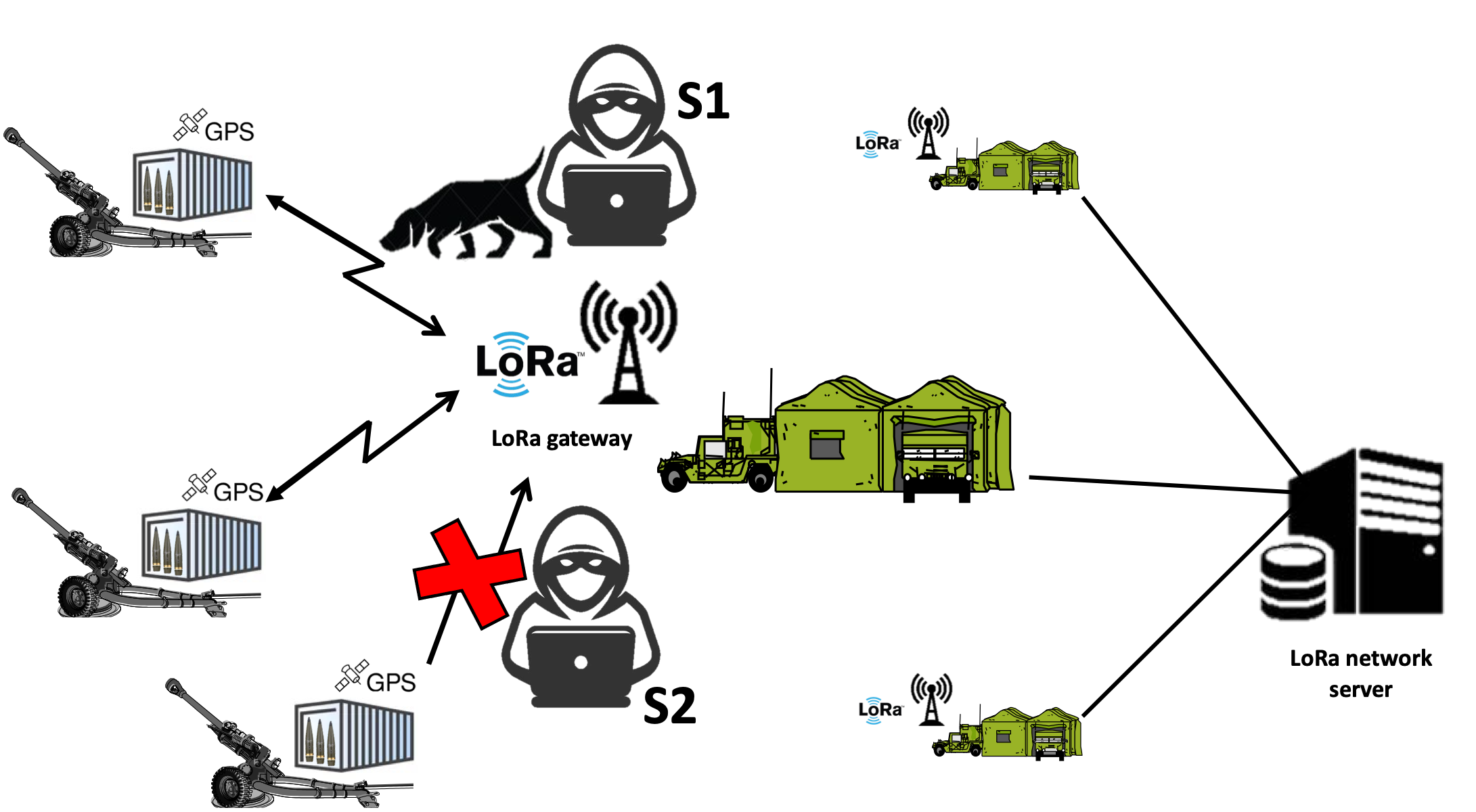}
    \caption{Attack scenarios}
    \label{fig:galaxy}
\end{figure}

We wanted to set up a simple scenario that could be accessible to all skill levels and applicable to both civil technology and military contexts. To achieve this, we created a simple targeted LoRaWAN network: one device, one gateway and a community network server operating under the release 1.0.x.  Our implementation was as follows: the device responsible for tracking the GNSS position of the simulated military equipment was a Heltech card (CubeCell GPS-6502) compatible with Arduino development environment combining a LoRaWAN transmission module, itself associated with a GPS positioning module. This device communicated with a gateway named Dragino LPS8 controlled by The Things Network (TTN) software. \\

In the first scenario  we set up the target LoRaWAN network, then sniff exchanged messages between the device and the gateway. We used the Dragino LPS8 and TTN software, as well as a HackRF software-defined radio. Then we analyzed metadata by reading LoRaWAN frame. For scenario 2, we send back the frames received to the target gateway. We first investigated the possibility of using Dragino LPS8 and TTN software to create the eavesdropping network. Once this network was functional, we implemented the attack network with a HackRF software-defined radio (see Fig. 2). 

\begin{figure}
    \centering
    \includegraphics[width=0.9\linewidth]{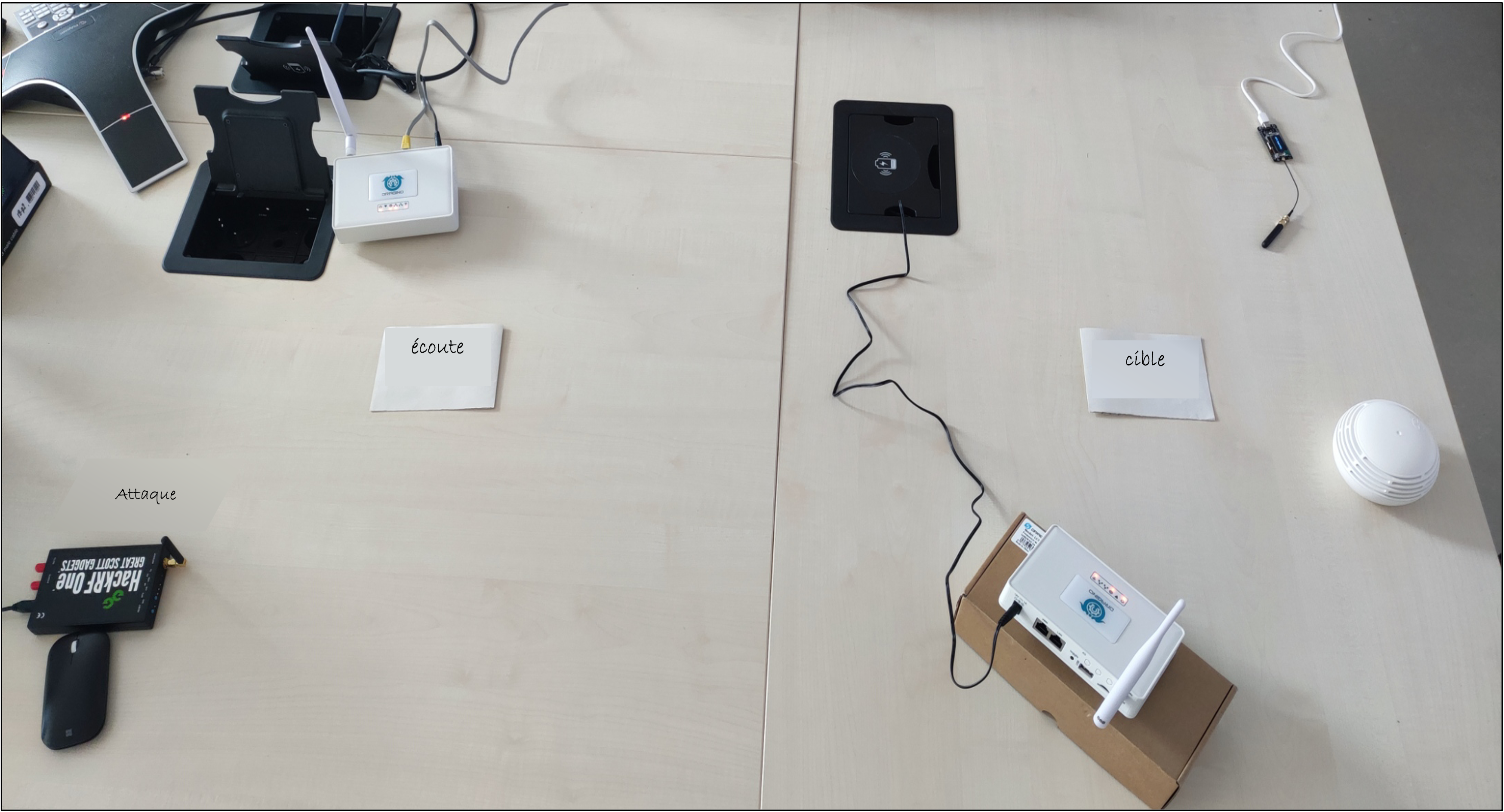}
    \caption{photo of the simulation network}
    \label{fig:enter-label}
\end{figure}

\section{Implementation and results}


\textbf{Sniffing}: during the study of the LoRaWAN protocol, we identify that the procedure of enrolling a device to the network use the "join request", which is transmitted unencrypted. In addition, a LoRaWAN device addresses all gateways in the coverage area that can receive messages and process them. "Join requests" message is therefore accessible to anyone listening to LoRaWAN frequencies and spreading factors. 

So, we intercept the "join request" frame of GPS-equipped device, which, combined with a first TTN analysis, allowing access to the following characteristics: deveui (unique identifier of the LoRaWAN object), devnonce (anti-replay counter), data rate (spreading factor, bandwidth, coding rate), frequency: 868.3 Mhz, timestamp: time of reception of the frame,  RSSI, frequency offset and signal to noise. \par 

It is then a question of deducing information of interest; the deveui can provide information about the manufacturer. In our experience it is an HT-MO1 model of the Heltech brand which mainly makes objects for localization purposes. In addition, RSSI, frequency offset, and signal to noise elements can provide distance information. In our study, we use an open-source software that aggregates this data to estimate the distance between gateway and node but more advanced solutions exist \cite{6}. Moreover, the variation in the RSSI intercepted suggests that the device is moving. It could therefore be a piece of equipment that sends the positions of the asset to which it is associated. However, it is important to be cautious about these conclusions, which would need to be confirmed and verified by other sensors. \\

\textbf{Replay attack}: our objective here is to test different types of replay attacks and to observe the consequences of rejection of the device by the TTN, and therefore its "deauthentication". In our scenario, we implement two replay attacks, the first one was about joining requests with identical devnonce, and the second on the replay of uplink packets (encrypted with an identical FCnt), as if the attacked node was rebroadcasting a previously sent packets.\par

To carry out these attacks, we use the native acquisition function of the HackRf to capture the LoRaWAN frames of interest on the radio interface. Then, we re transmit these frames by adjusting the amplification and the gain in dB.

Regarding the first attack, we observe the results as interpreted by the TTN server (fig 3). We see the different "join requests" issued by our replay platform. However, since the legitimate object is already connected, i.e. the devnonce already known, our hacker "join request" is not accepted by the server. A change of scale would undoubtedly cause the deauthentication of the object. \par

\begin{figure}[htp]
    \centering
    \includegraphics[width=0.8\linewidth]{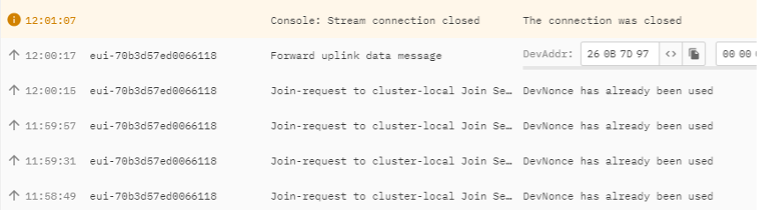}
    \caption{1st replay attacks observations}
    \label{fig:enter-label}
\end{figure}

For the second attack, we test different gateways (local or web) and note that the LNS drops packets with the error message "FCnt has already been used". This is because when a LoRaWAN node sends a packet, it includes a frame counter (FCnt) that is incremented with each send. The LoRaWAN server uses it to identify unique packets and avoid duplicates. If the LoRaWAN server receives a packet with a frame counter that is identical to a previously received packet, it discards the packet. 

\section{Conclusion}
In this study, we implemented eavesdropping and replay attacks scenarios in a military context using the LoRaWAN network. The results of our experimentation show the deauthentication of devices on LNS platforms other than TTN. This indicates the importance of offensive security to validate the use of such technology in critical context. As a future work, we can make more experimentation by variation of distance and time, as well as the number of proprietary sensors in order to generate another attack like the distributed denial of service.



\end{document}